\documentclass[useAMS,usenatbib,epsfig]{mn2e}
\newif\ifAMStwofonts
\AMStwofontstrue
\usepackage{amsmath,amsfonts,epsfig,natbib}

\def\reff@jnl#1{{\rm#1\/}}
\def\aj{\reff@jnl{AJ}}                  
\def\araa{\reff@jnl{ARA\&A}}            
\def\apj{\reff@jnl{ApJ}}                
\def\apjl{\reff@jnl{ApJ}}               
\def\apjs{\reff@jnl{ApJS}}              
\def\ao{\reff@jnl{Appl.Optics}}         
\def\apss{\reff@jnl{Ap\&SS}}            
\def\aap{\reff@jnl{A\&A}}               
\def\aapr{\reff@jnl{A\&A~Rev.}}         
\def\aaps{\reff@jnl{A\&AS}}             
\def\azh{\reff@jnl{AZh}}                
\def\baas{\reff@jnl{BAAS}}              
\def\gca{\reff@jnl{GeCoA}}              
\def\jrasc{\reff@jnl{JRASC}}            
\def\memras{\reff@jnl{MmRAS}}           
\def\mnras{\reff@jnl{MNRAS}}            
\def\pra{\reff@jnl{Phys.Rev.A}}         
\def\prb{\reff@jnl{Phys.Rev.B}}         
\def\prc{\reff@jnl{Phys.Rev.C}}         
\def\prd{\reff@jnl{Phys.Rev.D}}         
\def\prl{\reff@jnl{Phys.Rev.Lett}}      
\def\pasp{\reff@jnl{PASP}}              
\def\pasj{\reff@jnl{PASJ}}              
\def\qjras{\reff@jnl{QJRAS}}            
\def\skytel{\reff@jnl{S\&T}}            
\def\solphys{\reff@jnl{Solar~Phys.}}    
\def\sovast{\reff@jnl{Soviet~Ast.}}     
\def\ssr{\reff@jnl{Space~Sci.Rev.}}     
\def\zap{\reff@jnl{ZAp}}                
\def\nat{\reff@jnl{Nature}}             

\title[Constraints on the WHIM SZ signal from WMAP and SPT]
{Constraints on the Sunyaev-Zel'dovich signal from the
Warm Hot Intergalactic Medium from WMAP and SPT data}

\author[R. G\'enova-Santos et al.] {Ricardo G\'enova-Santos,$^{1,2}\thanks{E-mail:
  rgs@iac.es}$ I. Suarez-Vel\'asquez,
$^3$ F. Atrio-Barandela,$^4$ \newauthor and J. P. M\"ucket$^3$\\
$^1$ Instituto de Astrofis\'{i}ca de Canarias, 38200 La Laguna, Tenerife, Canary Islands, Spain\\
$^2$ Departamento de Astrof\'{\i}sica, Universidad de La Laguna (ULL), 38206 La Laguna,
Tenerife, Spain\\
$^3$ Leibniz Institut f\"ur Astrophysik, 14482 Potsdam, Germany\\
$^4$ F{\'\i}sica Te\'orica, Universidad de Salamanca, 37008 Salamanca, Spain\\}
\date{Accepted Received In original form}
\pagerange{\pageref{firstpage}--\pageref{lastpage}}
\pubyear{2013}
\begin{document}

\label{firstpage}
\maketitle

\begin{abstract}
The fraction of ionized gas in the Warm Hot Intergalactic Medium
induces temperature anisotropies on the Cosmic Microwave Background similar
to those of clusters of galaxies. The Sunyaev-Zel'dovich anisotropies due
to these low density, weakly non-linear, baryon filaments can not be distinguished
from that of clusters using frequency information, but they can be separated
since their angular scales are very different. To determine the relative
contribution of the WHIM SZ signal to the radiation power spectrum of
temperature anisotropies, we explore the parameter space of the concordance
$\Lambda$-Cold Dark Matter model using Monte Carlo Markov Chains and the
Wilkinson Microwave Anisotropy Probe 7yr and South Pole Telescope data.
We find marginal evidence of a contribution by diffuse gas, 
with amplitudes of $A_{\rm WHIM}=10-20~\mu$K$^2$, but the results are also 
compatible with a null contribution from the WHIM, allowing to set an upper limit 
of $A_{\rm WHIM}<43~\mu$K$^2$ ($95.4$\%~C.L.). The signal produced by galaxy 
clusters remains at $A_{\rm CL}=4.5~\mu$K$^2$, a value similar to what is 
obtained when no WHIM is included. From the measured 
WHIM amplitude we constrain the temperature-density phase diagram of 
the diffuse gas, and find it to be compatible with numerical simulations.
The corresponding baryon fraction in the WHIM varies 
from 0.43 to 0.47, depending on model parameters.
The forthcoming Planck data could 
set tighter constraints on the temperature-density relation.
\end{abstract}


\section{Introduction}

At low redshifts, close to half of the baryons in the Universe have yet 
to be identified (Fukugita \& Peebles, 2004). Numerical simulations
suggest the existence of a Warm-Hot Intergalactic Medium (WHIM) phase 
of mildly non-linear structures 
(Cen \& Ostriker, 1999, 2006; Dav\'e et al. 2004; Smith et al. 2011; 
Shull, Smith \& Danforth, 2012). Most of the efforts to detect the WHIM 
has concentrated in the identification of X-ray absorbers, and 
around half of the WHIM baryons could have been identified 
through this method, leaving the remaining missing baryon 
fraction at $\sim 29\pm 13\%$ (Shull et al. 2012). However, 
many of these detections at $z>0$ are either of low statistical 
significance or controversial. For instance, the X-ray detections  
by Nicastro et al. (2005) have not been confirmed by later studies
(Kaastra et al. 2006, Yao et al. 2012). 
As an observational alternative, Atrio-Barandela \& 
M\"ucket (2006) suggested that the WHIM could be detected through the 
temperature anisotropies generated on the Cosmic Microwave Background (CMB).
Due to thermal and kinetic motions of ionized gas, Compton scattering of 
CMB photons by free electrons induces secondary temperature anisotropies that
were first described by Sunyaev \& Zeldovich (1970, 1972, hereafter SZ, 
Birkinshaw, 1999). The thermal 
SZ (TSZ) temperature anisotropies have a distinctive frequency dependence, 
different from other foregrounds, and have been detected in the direction of 
many known clusters (see e.g. Planck Collaboration, 2011).  
The detection of the TSZ signal associated with 
the less-dense WHIM inter-cluster filaments is rather more challenging owing 
to its small amplitude. Hern\'andez-Monteagudo, G\'enova-Santos \&
Atrio-Barandela (2004) attempted to measure this signal though cross-correlations 
between CMB maps and templates of the density field constructed from galaxy 
catalogues (see also Suarez-Vel\'asquez et al. 2013a).
G\'enova-Santos et al. (2008) detected a temperature decrement towards the 
Corona Borealis supercluster, in a position with no known clusters but with an 
overpopulation of galaxies (Padilla-Torres et al. 2009), but could not confirm 
it was associated to WHIM gas. More recently, an unambiguous measurement 
of a hot and diffuse gas component outside the virial 
regions of clusters was obtained by combining X-ray and CMB data 
from the ROSAT and Planck satellites on the cluster pair A399-A401 
(Planck Collaboration, 2012). The other SZ component, the KSZ effect, is 
more difficult to detect even towards galaxy clusters since it has the 
same frequency dependence as the intrinsic CMB signal. Only the average KSZ 
effect due to large samples of clusters (Kashlinsky et al. 2008, 2010) or galaxies 
(Hand et al. 2012) has been detected. 

Frequency information can not be used to distinguish the TSZ anisotropy
due to clusters and WHIM filaments, but they could be separated using
the different scale of their respective anisotropies.
The cluster TSZ contribution is maximum at 
$\ell_{max}\sim 3000$ (Atrio-Barandela \& M\"ucket, 1999) while the 
thermal and kinematic SZ due to the WHIM is in the range 
$\ell_{max}\simeq 200-500$, depending on model parameters (Atrio-Barandela 
\& M\"ucket 2006, Atrio-Barandela, M\"ucket \& G\'enova-Santos 2008,
Suarez-Vel\'asquez, M\"ucket \& Atrio-Barandela 2013b). 
To successfully separate the WHIM from the cluster SZ contribution,
the latter needs to be well characterized.  The analysis of
the Wilkinson Microwave Anisotropy Probe (WMAP) 7 year data confirmed that
the SZ power spectrum due to the unresolved cluster population was a 
factor $0.45\pm 0.06$ the theoretical predictions and estimates based 
on numerical simulations (Komatsu et al 2011). This lower than expected 
amplitude agreed with the independent results from the South Pole Telescope 
(SPT, Keisler et al 2011), suggesting that the current models of the 
intracluster medium overestimates the gas pressure relative to X-ray and 
CMB observations (Komatsu et al. 2011). 
An alternative explanation relies on the fact that 
the TSZ power spectrum depends on the amplitude of the matter density fluctuations, 
$\sigma_8$, as  $C_\ell^{\rm CL}\propto(\sigma_8)^7$ (Komatsu \& Kitayama 1999).
If the measured value $\sigma_8=0.801\pm 0.030$ (Larson et al. 2011) were 
overestimated, so it would be the theoretical prediction for the TSZ 
amplitude. A lower $\sigma_8$ would also result in a lower amplitude of the 
primordial CMB anisotropies at low $\ell$, which then could accommodate 
a WHIM component to compensate for this power deficit. 
In G\'enova-Santos (2009) we showed that a 2-3\%
variation of the concordance $\Lambda$CDM model value of $\sigma_8$
allowed a contribution of the WHIM of amplitude 
$\ell(\ell+1)C_\ell/2\pi\simeq 100~\mu$K$^2$ at $\ell\sim 200$.

In the present paper we first analyze the combined WMAP 7yr and SPT data, 
the best available data set to date, constraining the power spectrum out to 
$\ell=3000$, in order to search for any WHIM SZ contribution 
to the CMB radiation power spectrum. 
Due to the difference in scale, the WHIM and cluster anisotropies
can be separated through their power spectrum.
Since WMAP probes scales below $\ell\le 1000$ it is mostly
sensitive to the WHIM. SPT data, that probes $\ell\ge 600$, can constrain better 
the cluster component. By combining both data sets, we could separate
both contributions to TSZ anisotropies. Second, we study if the WHIM component 
brings the cluster contribution in agreement with the theoretical and numerical 
expectations. In Sec.~2 we describe the WHIM model and its parameters;
in Sec.~3 we describe the power spectrum analysis implemented in Monte Carlo 
Markov Chains (MCMC); in Sec.~4 we discuss our main results and in Sec.~5 we 
summarize our conclusions.

\section{Temperature anisotropies generated by the Warm Hot Intergalactic Medium}

In our model, we define the WHIM as shock-heated intergalactic gas at density 
contrasts in the range $\delta_B=[1, 100]$ and temperatures $10^5$K$ < T < 
10^7$K. We take a rather low upper limit for $\delta_B$ in order to clearly distinguish  
the WHIM in the less dense inter-cluster filaments that form the cosmic web from the 
gas surrounding clusters of galaxies. The WHIM phase of our model encompasses 
the fraction of the missing baryons with $\delta_B\leq 100$. 
Due to the low upper limit of $\delta_B$ chosen by us, 
the contributions of the WHIM to the CMB anisotropy spectrum in our model 
must be considered as a lower limit to the true anisotropy.

To compute the TSZ power spectrum of this WHIM phase, we assume that 
at any given point the probability of having a filament with a baryon 
density contrast $\delta_B>1$ is given by the log-normal probability density 
function (PDF, Atrio-Barandela \& M\"ucket 2006). The log-normal PDF was 
introduced by Coles \& Jones (1991) to describe the non-linear distribution 
of matter in the Universe. It has been applied to the column density distribution 
for neutral hydrogen in the intergalactic medium (Choudhury et al. 2001) and 
it has been found to describe very well the matter statistics at scales larger 
than $7h^{-1}$Mpc (Kitaura et al. 2009). In the log-normal model,
the power spectrum of the baryon density inhomogeneities follows the 
dark matter distribution above a linear scale $L_0$. Below this
cut-off scale, baryon density perturbations are smoothed out due to physical 
effects like Jeans dissipation or shock heating.  At the physical 
conditions present in the WHIM, shock heating is the dominating process
and $L_0(z)$ is the scale at 
which the linear peculiar velocity $ v_p(\bf x,z)$ is equal or 
larger than the sound speed $c_s(z)$ of the baryon fluid at redshift $z$
(for details, see Suarez-Vel\'asquez et al. 2013b).
The sound speed is given by the mean temperature of the Intergalactic
Medium (IGM) as $c_s=(k_{\rm B}T_{\rm IGM}(z)/m_p)^{1/2}$, where $m_p$ is the proton
mass. At every redshift, the IGM temperature varies with
location, but the average value, $T_{\rm IGM}$, determines the 
smallest baryon density perturbation that survives shock heating.
The evolution of the IGM temperature is mostly determined by the UV background. 
The mean IGM temperature variation with redshift is small and we
approximate it by $\log_{10}(T_{\rm IGM}/10^3{\rm K})=(A+0.1(1+z))$ (Theuns et al. 2002).
At redshifts $z\le 3$ the temperature varies in 
the range $T_{\rm IGM}=10^{3.6}-10^{4.0}$~K (Tittley \& Meiksin, 2007);
consequently, $A=0.5-0.9$. At redshifts $z>3$ the WHIM does not generate 
significant temperature anisotropies and its contribution can be ignored.
When $T_{\rm IGM}$ increases, more baryon fluctuations are erased and the
size of baryon filaments increases, subtending a larger angle and giving 
rise to CMB temperature anisotropies at larger angular scales.
The corresponding scale at $z=0$ is
$L_0\simeq 1.7(T_{\rm IGM}/10^{4.0}K)^{1/2}h^{-1}$Mpc. 

\begin{figure}
\centering
\includegraphics[width=\columnwidth]{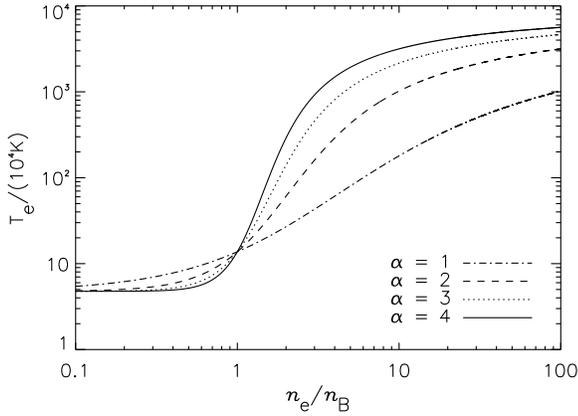}
\caption{\small Different fits of the phase diagram
of the hydrodynamical simulations of Kang et al (2005). From bottom to top the 
dotted, dashed, dot-dashed and solid lines correspond to values 1,2,3 and 4
of the $\alpha$ parameter given in eq.~(\ref{eq:fit}).}
\label{fig:temperature}
\end{figure}

To compute the TSZ temperature anisotropies, it is necessary to determine
the number of free electrons in each filament and their average temperature.
The number density of electrons $n_e$ can be obtained by assuming equilibrium 
between ionization (collisional and photo-ionization) and recombination. At 
temperatures $T=10^5-10^7$~K and density contrasts $\delta_B\le 100$, the 
gas can be considered fully ionized. Whereas in Atrio-Barandela \& M\"ucket (2006) 
we used a polytropic equation of state, in Suarez-Vel\'asquez
et al. (2013b) we used fits of the phase diagrams obtained in various
hydrodynamical simulations to take into account the effect of shock heating 
in the evolution of the WHIM.  We used 
\begin{equation}
\log_{10}\left(\frac{T_{\rm e}(\xi)}{10^8{\rm K}}\right) =
-\frac{2}{\log_{10}(4+\xi^{\alpha+1/\xi})}~~,
\label{eq:fit}
\end{equation}
since this numerical fit reproduces well the phase diagram of Kang et al. (2005). 
In this expression, $\xi=n_e/\bar{n}_{B}\simeq\delta_B+1$ is the electron 
density in units of the mean baryon density. The phase diagrams derived from the 
simulations are not simple linear relations of the type $T=T(\xi)$, but 
have a large scatter. The $\alpha$ parameter of the previous equation is introduced 
to model this uncertaintly. We consider an interval wide enough, from $\alpha = 1$ 
to $\alpha = 4$, in order to properly cover all the possible variations found in 
the simulations. In Fig.~\ref{fig:temperature} we represent the resulting 
equations of state for $\alpha=(1,2,3,4)$ in decreasing order from top to bottom.
\begin{figure*}
\centering
\includegraphics[width=15cm]{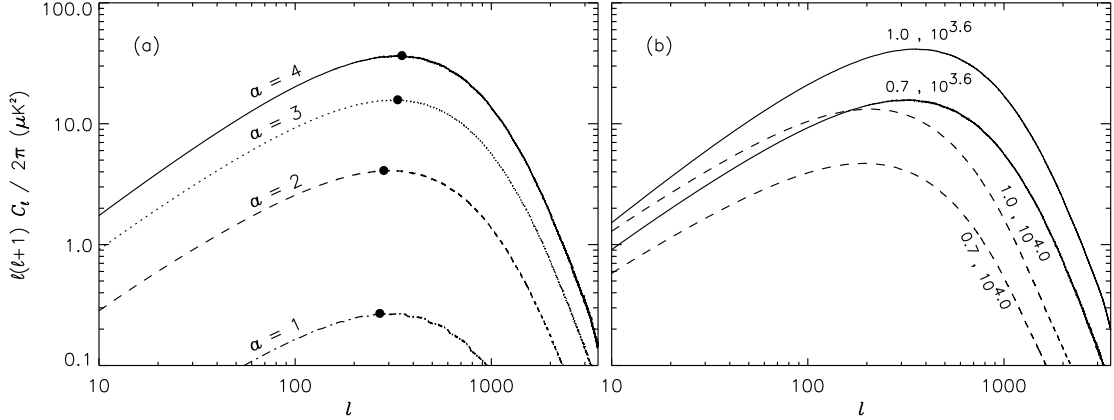}
\caption{Radiation power spectra for different WHIM model parameters.
(a) Variation with the $\alpha$ parameter given in eq.~\ref{eq:fit}.
From bottom to top the dot-dashed, dashed,
dotted and solid lines correspond  respectively to $\alpha=1,2,3,4$, 
with $T_{\rm IGM}=10^{3.6}$~K and $\sigma_8$=0.7. All other parameters
are fixed to the values of the concordance $\Lambda$CDM model.
The maxima, indicated by filled circles, are in the range $\ell=[270,350]$. 
(b) Variation with the IGM temperature and $\sigma_8$ for $\alpha=3$.
Solid and dashed lines correspond to $T_{\rm IGM}=10^{3.6},10^{4.0}$~K;
for each pair, the top and bottom lines correspond to $\sigma_8=0.7,1.0$,
respectively. }
\label{fig:cl}
\end{figure*}

In our model, the correlation function of the spatial variations of the electron
pressure along two lines of sight separated by an angle $\theta$ is 
\begin{equation}
C(\theta)= \int_0^{z_f}\int_0^{z_f}\langle 
S(\hat{x}_1,z_1)S(\hat{x}_2,z_2)\rangle dz_1 dz_2 .
\label{cfull}
\end{equation}
In this expression $S=G(\nu)(k_{\rm B}\sigma_T/m_ec^2) n_eT_e(dl_i/dz_i)$ 
with $i=(1,2)$. Integrations are carried out along the line of sight $l$. 
For each direction, the TSZ WHIM temperature anisotropy is 
$\Delta T=y_cG(\nu)$, with $y_c=k_{\rm B}\sigma_T/m_ec^2\int T_e n_e dl$ 
the integrated Comptonization parameter, $n_e$ the electron density,
$T_e$ the electron temperature, $m_ec^2$ the electron 
annihilation temperature, $k_{\rm B}$ the Boltzmann constant, $\sigma_T$ Thomson 
cross section, $dl$ the line element, $dz$ its corresponding redshift interval 
and $G(\nu)=(x\coth(x/2)-4)$ the frequency dependence of the TSZ effect with
$x=h\nu/kT_0$ the reduced frequency and $T_0$ the CMB temperature.
Expanding eq.~(\ref{cfull}) in Legendre polynomials gives the WHIM 
contribution to the power spectrum of CMB temperature anisotropies
\begin{equation}
C_\ell=2\pi\int_{-1}^{+1}C(\theta)P_\ell(\cos\theta)d\cos\theta
\label{cl}
\end{equation}
where $P_\ell$ is the Legendre polynomial of multipole $\ell$.

The radiation power spectrum depends on the underlying cosmological
model and the parameters describing the physical state of the WHIM, as 
shown in Fig.~\ref{fig:cl}. The $\Lambda$CDM cosmological
parameters have been fixed to the best fit values of WMAP 7yr data.
In Fig.~\ref{fig:cl}a we represent the spectra for different fits 
of the equation of state. The IGM temperature, that parametrizes
the cut-off length of the baryon power spectrum $L_0$, is $T_{\rm IGM}=10^{3.6}$~K, 
while the amplitude of matter fluctuations is $\sigma_8=0.7$.
From bottom to top the fit parameter of the equation of state is $\alpha=1,2,3,4$.
At densities $\xi\ge 2$ the temperature of the free electrons 
in the WHIM increases with increasing $\alpha$ and the resulting
temperature anisotropies also increase. 
The filled circles show the position of the maxima, that in this case vary
in the range $\ell=_{\rm max}[270,350]$. The variation is smaller for larger $\sigma_8$.
These maxima correspond to angular scales
in the range $\simeq 40-50$~arcmin, much larger than $\sim 7$~arcmin,
the scale of the power spectrum of the unresolved 0.5-2~keV
Cosmic X-ray Background due to both clusters and diffuse gas
(Roncarelli et al. 2012, Cappelluti et al. 2012). The difference
in scale between the power spectrum of TSZ anisotropies due
to clusters and due to the WHIM is not in contradiction with 
our result. The X-ray power spectra includes the contribution
of diffuse gas from denser environments ($\delta\le 10^3$); 
excluding those high dense regions shifts the
power of the WHIM anisotropies scales larger than those of clusters. 
Further, since the TSZ distortion is proportional to the electron pressure, 
$n_eT_e$ while X-ray emissivity is proportional to $n_e^2T_e^{1/2}$ (Birkinshaw 1999),
the TSZ anisotropy will be more extended than the X-ray emission of the
WHIM, as it occurs with clusters (Atrio-Barandela et al 2008).

In Fig.~\ref{fig:cl}b
we represent the variation with $T_{\rm IGM}$ and $\sigma_8$. The solid
and dashed lines correspond to $T_{\rm IGM}=10^{3.6},10^{4.0}$~K, 
respectively. For each pair, the top and bottom lines correspond to
$\sigma_8=1.0$ and $0.7$. The fit parameter of the equation of state is fixed at
$\alpha=3$. This figure shows that the TSZ WHIM
temperature anisotropies grow with increasing $\sigma_8$ and decreasing
$T_{\rm IGM}$. Lowering the cut-off length results in larger anisotropies
since more baryon density perturbations at small scales survive shock
heating. As a result, the average angular size of baryon filaments decreases
and the power spectrum is shifted to small angular scales.

\section{Markov Chain Monte Carlo analysis of WMAP and SPT}

The shape of the SZ power spectrum, $\ell(\ell+1)C_\ell/2\pi$, is very 
similar for both clusters and the WHIM; it presents a single maximum
at $\ell_{max}$, determined by the average angular size of clusters 
and baryon filaments. The main difference is the location of their maxima.
As indicated in the introduction, it is this difference on 
angular scale that can be used to separate the contribution of
the WHIM from that of clusters. In both cases, the overall shape
of the spectrum is very weakly dependent on other model parameters. 
For instance, due to the uncertainties on the number density and
pressure profile and on their scaling with cluster mass and 
redshift, Keisler et al. (2011) used a fixed spectral shape, derived from 
numerical simulations that contains both the TSZ and KSZ components, and 
fit its amplitude to the data. If the radiation power spectrum is
written as $C_\ell^{\rm SZ}=A_{\rm SZ} G(\nu)^2 f(\ell)$, then $f(\ell)$ is
kept constant and only $A_{\rm SZ}$ is derived from the data. 
To search for any WHIM SZ component in addition to the 
SZ due to clusters we will explore the parameter space of the standard
$\Lambda$CDM model by fitting the theoretically predicted radiation power 
spectra to WMAP and SPT data including WHIM and cluster contributions. 
We use the {\sc cosmomc} package (Lewis \& Bridle 2002), which implements a MCMC 
method that performs parameter estimation using a Bayesian approach. 
We modified the Keisler et al. 
(2011)\footnote{Likelihood code downloaded from\\ 
{\tt http://pole.uchicago.edu/public/data/keisler11/\#Likelihood}} version by
adding at each step of the chain the precomputed WHIM SZ component to the 
theoretical power spectrum computed with {\sc camb} (Lewis et al. 2000). 

We constructed a three-dimensional grid of WHIM models for values of 
$\sigma_8=0.7$ and 1.0, $\alpha = 1,2,3,4$ and $\log(T_{\rm IGM}/10^3K)=A+0.1$ 
with $A=[0.3,1.1]$ varying in units of $0.1$. Firstly, 
the power spectra for any given value of these parameters were interpolated 
between the pre-computed spectra corresponding to the 
nearest adjacent values. The interpolation was linear in $A$ and logarithmic
in $\alpha,\sigma_8$. For the latter, the spectra scales as
$C_\ell^{\rm WHIM}\propto \sigma_8^{2.6}(\Omega_{\rm B}h)^2$
(Suarez-Vel\'asquez et al. 2013b).
Even if the interpolation procedure was accurate (errors below 1\%), the chains were 
slowly convergent and degenerate with respect to $T_{\rm IGM}$. For this reason, 
we removed the interpolation on the temperature and fixed $T_{\rm IGM}$.
Fig.~\ref{fig:cl} illustrates that varying $\alpha$ (Fig.~\ref{fig:cl}a) 
and $\sigma_8$ (Fig.~\ref{fig:cl}b) changes the amplitude but does not 
significantly modify the shape of the radiation power spectrum. 
To see the effect of the location of the maximum in 
our final results, we considered two fixed WHIM power spectra with 
maxima located at $\ell_{max}=330$ and $210$. Neglecting the small
variations due to the different values of $\alpha$, this maxima correspond to 
$T_{\rm IGM}=10^{3.6}$ and $10^{4.0}$~K, respectively.
Therefore, similarly as it is done for the cluster contribution,
in our MCMC's we input these two spectra and
fit their amplitude ($A_{\rm WHIM}$) to the data. 

We fit the data, through the combined WMAP and SPT likelihoods, to a 
concordance $\Lambda$CDM model, defined by a spatially flat Universe with 
cold dark matter (CDM), baryons, and a cosmological constant $\Lambda$. 
Following Keisler et al. (2011) we added the combined galaxy clusters 
TSZ+KSZ amplitude ($A_{\rm CL}$) and two amplitudes associated with 
the contributions from Poisson ($A_{\rm PS}^{\rm Poisson}$) and clustered 
($A_{\rm PS}^{\rm Clustered}$) distributed point sources in addition to
the amplitude of the WHIM TSZ signal ($A_{\rm WHIM}$).
The WHIM and cluster power spectra were normalized to unity at $\ell=300$ 
and $\ell=3000$, respectively. Then, to take into account the
frequency dependence of the TSZ effect, their amplitudes $A_{\rm WHIM}$ and 
$A_{\rm CL}$ were scaled to the average frequency of WMAP V and W bands
($\langle G(\nu)^2\rangle=2.861$) and of the three SPT bands 
($\langle G(\nu)^2\rangle=1.107$). Note that in our analysis we are 
neglecting the less important KSZ component of the WHIM. Appart from being 
fainter, this is justfied because the TSZ and KSZ WHIM power spectra have very 
similar shapes, rendering a joint fit of both components very degenerate.

\begin{figure*}
\centering
\includegraphics[width=12cm]{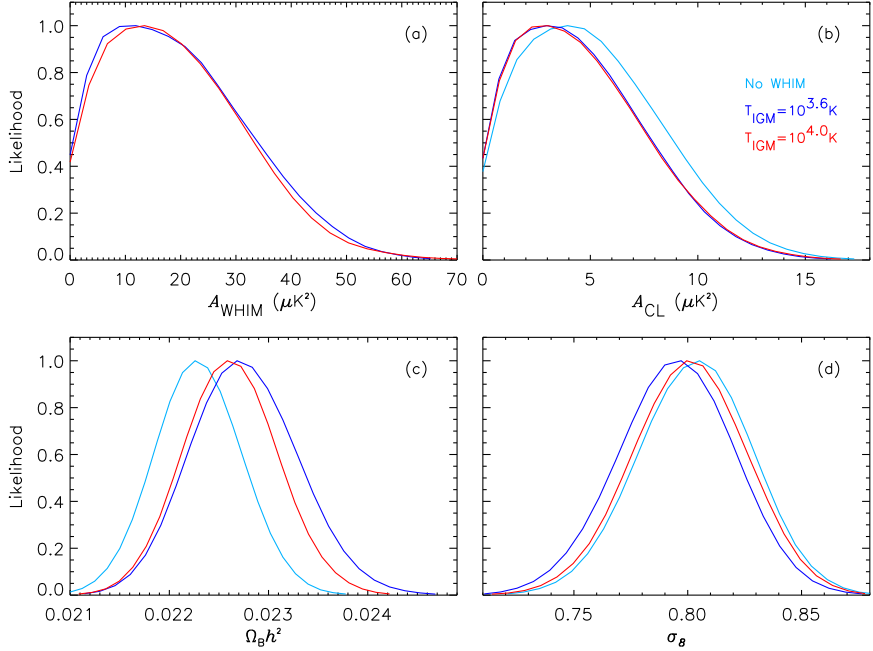}
\caption{\small Marginalized likelihoods for the WHIM TSZ amplitude at 
$\ell = 300$ ($A_{\rm WHIM}$), for the combined TSZ+KSZ amplitude at $\ell=3000$ 
from galaxy clusters ($A_{\rm CL}$), for the physical baryon density ($\Omega_{\rm B}h^2$), 
and for the amplitude of matter density fluctuations in scales of 8~h$^{-1}$~Mpc 
($\sigma_8$). We plot our results for the two IGM 
temperatures, $T_{\rm IGM}=10^{3.6}$~K (blue) and 10$^{4.0}$~K (red), 
as well as when no WHIM signal is introduced (cyan).}
\label{fig:like}
\end{figure*}

\begin{figure*}
\centering
\includegraphics[width=\textwidth]{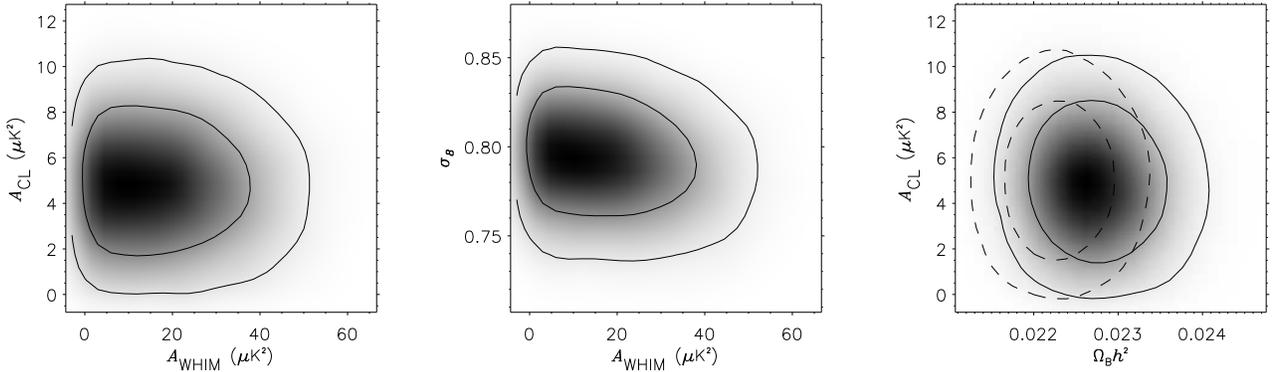}
\caption{\small Mean two-dimensional likelihoods for three combination of parameters, 
for $T_{\rm IGM}=10^{3.6}$~K. Solid lines represent the 
$1\sigma$ and $2\sigma$ confidence regions. Dashed lines in the
right panel correspond to the case with no WHIM component.}
\label{fig:like_2d}
\end{figure*}

We computed MCMCs in three different cases: to reproduce the analysis of Keisler 
et al. (2011) one run contained only the clusters SZ contribution; two 
other runs included both cluster and TSZ WHIM contributions, each with a 
fixed value of the IGM temperature, $T_{\rm IGM}=10^{3.6}$ and $10^{4.0}$~K. 
For each of these three cases, we ran eight independent chains with a total number 
of $\sim 350,000$ samples. The Gelman \& Rubin (1992) criterion showed
that all our chains had converged; in all cases the $R$ 
statistic was well below 1.2. For instance, $R\approx 1.01$ for 
$A_{\rm WHIM}$ and $\approx 1.003$ for the other parameters, when the WHIM 
component was included, and $R\approx 1.001$ when it was not.
Keisler et al. (2011) used a Gaussian 
prior in the SZ cluster amplitude, $A_{\rm CL}=5.5\pm 3.0~\mu$K$^2$, derived 
by Shirokoff et al. (2011) from an analysis of an earlier SPT data release. 
To determine the bias introduced by this prior on the WHIM amplitude, 
we ran our MCMCs with and without prior. No significant differences were 
found (except, of course, on the best-fit value of $A_{\rm CL}$ itself)
and henceforth we will quote results with no prior.

\section{Results and discussion}

In Fig.~\ref{fig:like} we represent the marginalized likelihood functions 
of the amplitude of the TSZ anisotropy generated by WHIM (Fig.~\ref{fig:like}a), 
TSZ and KSZ anisotropy generated by clusters (Fig.~\ref{fig:like}b), physical 
baryon density (Fig.~\ref{fig:like}c) and of the amplitude of matter density 
perturbations at $8h^{-1}$Mpc (Fig.~\ref{fig:like}d). For an easier comparison, 
all likelihoods were normalized to unity. As mentioned above, the results presented
do not include a prior on $A_{\rm CL}$. Blue and red lines correspond to 
$T_{\rm IGM}=10^{3.6}$ and $10^{4.0}$~K, while the cyan lines correspond
the results when no WHIM is included. Adding a WHIM component marginally 
decreases $A_{\rm CL}$ and $\sigma_8$. 
In Fig.~\ref{fig:like_2d} we plot the two-dimensional likelihoods for different
pairs of parameters. The figure shows that $A_{\rm WHIM}$ is 
degenerate with respect to other parameters, especially with 
respect to $A_{\rm CL}$ and $\sigma_8$, and that it remains 
compatible with zero at the 2-$\sigma$ level. The 1-$\sigma$ contours 
of the $A_{\rm WHIM}-\sigma_8$ plot appear slightly curved downwards for increasing 
$A_{\rm WHIM}$, reflecting the (marginal) decrement of $\sigma_8$ when 
we include the WHIM TSZ anisotropy. Finally, the right panel shows 
how $\Omega_{\rm B}h^2$ decreases when there is no WHIM contribution
while $A_{\rm CL}$ remains almost unaffected. The dashed lines show
the contours when no WHIM anisotropy is included.

\begin{figure*}
\centering
\includegraphics[scale=0.6]{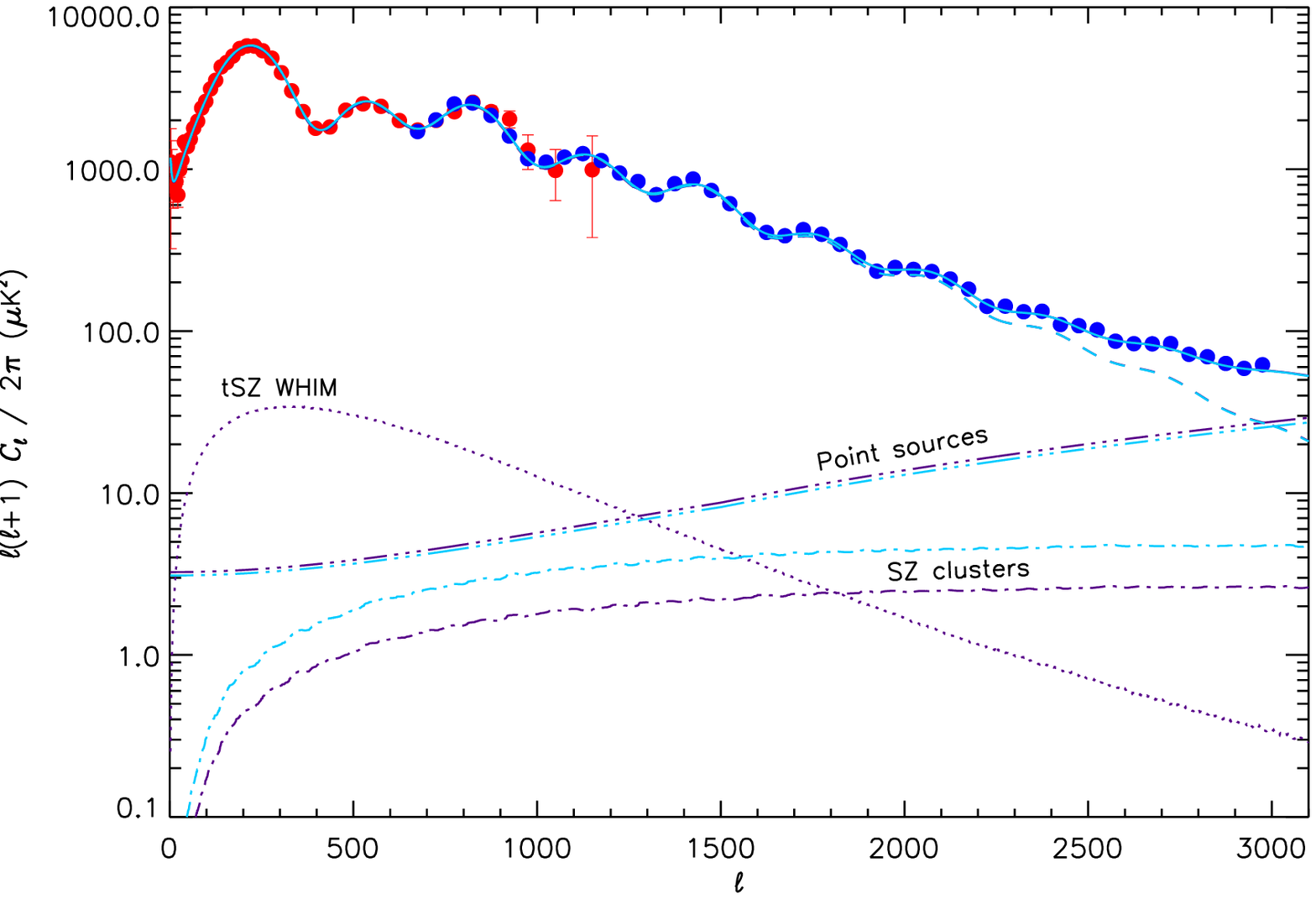}
\caption{\small Experimental power spectra measured by WMAP7 (red dots) 
and SPT (blue dots) compared with our fitted models. Dashed, dotted, 
dashed-dotted and dashed-triple-dotted lines correspond respectively to 
the primordial CMB, WHIM tSZ, combined tSZ and kSZ signals from clusters, 
and point sources (combing the terms coming from Poisson and 
clustered distributed sources). Solid lines represent the sum 
of all these components. We compare our results including the WHIM 
TSZ component (magenta curves) with the case when no WHIM 
component is included (cyan curves). This plot corresponds to 
the case of $T_{\rm IGM}=10^{3.6}$~K, and shows no significant differences 
with respect to the case of $T_{\rm IGM}=10^{4.0}$~K.}
\label{fig:power_spectra}
\end{figure*}

In Fig.~\ref{fig:power_spectra} we compare the measured 
WMAP 7yr (red filled circles) and SPT (blue filled circles) with our best-fit 
models in two cases (a) with a WHIM anisotropy with  $T_{\rm IGM}=10^{3.6}$~K and 
(b) without a WHIM component, represented by (indiscernible) solid lines.
We also plot the different components: primordial CMB (dashed lines), TSZ from 
WHIM (dotted line), SZ from galaxy clusters (dashed-dotted lines) and emission 
from point sources (both Poisson and clustered terms; dashed-triple-dotted lines). 
For visual purposes, we have scaled the WHIM component to the WMAP 
frequencies and the cluster component to the SPT frequencies.

Table~1 shows the parameters that define our best-fit cosmological model for 
our three MCMC (without $A_{\rm CL}$ prior). These parameters are derived from the 
median of the marginalized posterior probability density function, and their confidence 
intervals encompass the 68\% of the probability around those points. 
The parameters derived when no WHIM component is included differ by  
less than 2\% from those of Keisler et al. (2011). In the three cases considered
by us, the amplitude of the SZ signal from clusters or from the WHIM 
is compatible with zero at the $2\sigma$ level. 
From the marginalized likelihood distributions we derive upper limits, at the 
$95.4$\% C.L., of $A_{\rm WHIM}<43.6~\mu$K$^2$ and $<42.7~\mu$K$^2$, for 
$T_{\rm IGM}=10^{3.6}$~K and $10^{4.0}$~K, respectively.
Taking the most-likeliy $A_{\rm WHIM}$ at face value, from the measured values of 
$\Omega_{\rm B}h^2$, $H_0$ and $\sigma_8$ given in the table, we can determine 
the value of $\alpha$ that reproduces the amplitude of $A_{\rm WHIM}$ for each
value of $T_{\rm IGM}$. By interpolating in our grid of models
we find $\alpha=2.46, 3.40$ for $T_{\rm IGM}=10^{3.6},10^{4.0}$~K, respectively. 

For each of these values of $\alpha$ we can compute the baryon content 
residing in the WHIM filaments with overdensities in the range 
$[\xi_1,\xi_2]$. To this aim, we use the formalism described in Suarez-Vel\'asquez et al. (2013b). 
The mass fraction is given by 
$M(\xi_1,\xi_2)=\bar{\xi}^{-1}\int_{\xi_1}^{\xi_2}\xi F(\xi)$, 
where  $\bar{\xi}$ is the mean of the distribution and $F(\xi)$
is the log-normal probability distribution function. The integration
range must be limited to those overdensities where the non-linear evolution is well 
described by our model. As a criteria, we took $\xi_2\le 100$ and 
$\xi_1=\xi_{\mathrm{median}}+\sigma$, with 
$\sigma$ the standard deviation of the 
log-normal distribution. To include only scales that undergo shock 
heating we also require as necessary condition that $\xi_1>2$ at all redshifts.
Since $\xi_1$ and $T_{\rm IGM}$ are redshift dependent, the fraction 
of baryons residing in the WHIM filaments is a decreasing function 
of $z$. For $\alpha=2.46, 3.40$ and $T_{\rm IGM}=10^{3.6},10^{4.0}$~K 
the derived baryon fractions at $z=0$ are respectively
$\Omega_{\rm B,WHIM}/\Omega_{\rm B} \approx 0.43$ and $0.47$.
Notice that even if the value of $\alpha$ is rather uncertain,
the baryon fraction varies little within the allowed parameter space.
Our result is slightly above the $0.29\pm 0.13$ fraction of
baryons that remain unidentified according to Shull et al. (2012).
The baryons detected through TSZ anisotropies will include baryons 
on low dense regions identified by other techniques, like some of 
the OVI absorption systems; after correcting the double accounting, 
the fraction of identified baryons would be close to unity.

In our chains we have fixed $T_{\rm IGM}$; then, including the WHIM contribution 
adds only one parameter, $A_{\rm WHIM}$ (that is related with $\alpha$), to the chain. 
The improvement in 
the quality of the fit is minimal: in the best case $\Delta\chi^2\simeq -0.4$. 
Therefore the data fails to provide any significant evidence of the existence 
of the WHIM. Not unexpectedly, when including a WHIM anisotropy, $A_{\rm CL}$
decreases by 10\%, enhancing the disagreement between the measured
cluster SZ anisotropies and the theoretical expectation.
However, the discrepancy is alleviated since $\sigma_8$ also decreases.
In the model with $T_{\rm IGM}=10^{3.6}$~K, the lower value of 
$\sigma_8$ reduces the theoretical SZ signal by 7\%
compensating the lower value of $A_{\rm CL}$ found. To make the cluster 
SZ signal compatible with the theoretical predictions we would need 
$\sigma_8\approx 0.73$, much smaller than the measured value.

\begin{table*}
\begin{center}
\begin{tabular}{l c c c}
\hline\hline
\noalign{\smallskip}
Parameter & No WHIM & WHIM $T_{\rm IGM}=10^{3.6}$~K & WHIM $T_{\rm IGM}=10^{4.0}$~K \\
\noalign{\smallskip}
\hline
\noalign{\smallskip}
$\Omega_{\rm B}h^2$ & $ 0.0223\pm 0.0004$ & $ 0.0228\pm 0.0005$ & $ 0.0226\pm 0.0005$\\
\noalign{\smallskip}
$\Omega_{\rm CDM} h^2$ & $ 0.110\pm 0.005$ & $ 0.111\pm 0.005$ & $ 0.111\pm  0.005$ \\
\noalign{\smallskip}
100$\theta$ & $ 1.041\pm 0.002$   & $ 1.042\pm 0.002$   & $ 1.042\pm 0.002$  \\
\noalign{\smallskip}
$\tau$ & $ 0.0857^{+ 0.0063}_{- 0.0070}$& $0.0864^{+ 0.0064}_{- 0.0072}$ & $0.0865^{+ 0.0064}_{- 0.0071}$ \\
\noalign{\smallskip}
$n_{\rm S}$& $ 0.964\pm 0.011$  & $ 0.962\pm 0.011$ & $ 0.967\pm 0.011$ \\
\noalign{\smallskip}
${\rm ln}(10^{10} A_{\rm S})$  & $ 3.19\pm 0.04$ & $ 3.18\pm 0.04$ & $ 3.17\pm 0.04$\\
\noalign{\smallskip}
$Y_{\rm He}$ & $ 0.2478\pm 0.0002$ & $ 0.2480\pm 0.0002$ & $ 0.2479\pm 0.0002$\\
\noalign{\smallskip}
$A_{\rm WHIM}$ &  & $19.36^{+13.37}_{-13.34}$   & $19.16^{+12.65}_{-12.91}$  \\
\noalign{\smallskip}
$A_{\rm CL}$&$5.15^{+ 3.34}_{- 3.37}$&$ 4.58^{+ 3.11}_{- 3.14}$&$4.60^{+ 3.12}_{- 3.14}$\\
\noalign{\smallskip}
$A_{\rm PS}^{\rm Poisson}$&$20.40^{+ 2.93}_{- 2.92}$&$20.23\pm 2.89$&$20.32^{+ 2.89}_{- 2.87}$\\
\noalign{\smallskip}
$A_{\rm PS}^{\rm Clustered}$&$ 5.11^{+ 2.22}_{- 2.28}$ & $ 5.03^{+ 2.21}_{- 2.25}$&$5.05^{+2.20}_{- 2.26}$\\
\noalign{\smallskip}
\hline
\noalign{\smallskip}
$\sigma_8$ & $ 0.803\pm 0.024$ & $ 0.795\pm 0.025$ & $ 0.800\pm 0.024$ \\
\noalign{\smallskip}
$\Omega_\Lambda$ & $ 0.735\pm 0.025$ & $ 0.738\pm 0.025$ & $ 0.736\pm 0.025$ \\
\noalign{\smallskip}
$\Omega_{\rm M}$ & $ 0.265\pm 0.025$ & $ 0.262\pm 0.025$ & $ 0.264\pm 0.025$ \\
\noalign{\smallskip}
$z_{\rm re}$ & $10.40^{+1.17}_{- 1.16}$ & $10.33\pm 1.18$ & $10.39^{+ 1.18}_{- 1.17}$ \\
\noalign{\smallskip}
$H_0$ & $71.00^{+ 2.13}_{- 2.14}$ & $71.53^{+ 2.19}_{- 2.20}$ & $71.27\pm 2.14$ \\
\noalign{\smallskip}
\hline
\noalign{\smallskip}
$\chi^2$ &3756.04 & 3755.84 &  3755.66  \\
\noalign{\smallskip}
\hline \hline
\end{tabular}
\end{center}
\normalsize
\medskip
\caption[tab:param_results]{Concordance $\Lambda$CDM best fit parameters 
for each MCMC. We show the results for the three cases that we have considered: 
no WHIM, and a WHIM component with $T_{\rm IGM}$ temperatures of $10^{3.6}$ 
and $10^{4.0}$~K, respectively. Parameters are derived from the median of 
the marginalized posterior probability density functions, while the confidence 
intervals are calculated from the 68\% area around those points. The last row 
shows the $\chi^2$ of these models. The amplitudes $A_{\rm WHIM}, A_{\rm CL}, 
A_{\rm PS}^{\rm Poisson}, A_{\rm PS}^{\rm Clustered}$ are given in
units of $\mu K^2$ and the Hubble constant $H_0$ in kms$^{-1}$Mpc$^{-1}$.}
\label{tab:param_results}
\end{table*}

The results of Table~1 can be understood taking into account the difference in
the angular scales 
probed by WMAP and SPT. The information on the radiation power spectrum provided 
by these instruments is complementary with each other. In WMAP, at low $\ell$'s 
errors are dominated by sampling variance and at high $\ell$'s by noise. By 
contrast, the radiation spectrum has been best measured by the SPT in the range 
$\ell\simeq 600-3000$. Then, while WMAP data is mostly sensitive to the range 
where the WHIM contribution is largest, around the first and second acoustic 
peaks, SPT data  constrain better the cluster TSZ contribution. The first 
acoustic peak occurs at $\ell\simeq 200$, its amplitude is 
$\ell(\ell+1)C_\ell/2\pi\simeq 6000~\mu$K$^2$ and scales approximately as 
$\sigma_8^2$. In $\Lambda$CDM if all parameters are held fixed but $\sigma_8$ 
decreases by 0.5\%, the amplitude of the first acoustic peak decreases 
by $\sim$1\%. i.e., there could be a WHIM as high as $60~\mu$K$^2$ at the maximum.
Since WMAP operates at Rayleigh-Jeans frequencies, this corresponds to an amplitude 
$A_{\rm WHIM}=15-20~\mu$K$^2$, compatible with the results
of Fig.~\ref{fig:like}. Even though the WHIM contribution falls for
$\ell\ge 1000$, the tail of the distribution overlaps with cluster
anisotropies and the SPT data would suppress the latter 
(see Fig.~\ref{fig:power_spectra}).

Since our results are compatible with zero WHIM
contribution, one could interpret them as an indication (1) 
that the data has not enough statistical power to identify the
WHIM contribution or (2) that the WHIM is not a strong contributor
to the baryon budget. While the estimated values of the total baryon 
mass fraction in groups and clusters are still lower than the latest 
CMB measurement of the same quantity (Giodini  et al. 2009), it seems
implausible that a better modelling of the gas physics could eventually  
solve the WHIM problem.  If clusters and groups store more baryons than 
presently believed, the discrepancy between the SZ amplitude of clusters 
measured by SPT and WMAP with the numerical expectations summarized in
Komatsu et al (2011) would be even more acute. What our results indicate 
is the difficulty to accommodate a WHIM component that brings the 
measured cluster TSZ signal in agreement with the theoretical predictions. 
The reason of this discrepancy might not be the measured value of $\sigma_8$ 
or $A_{\rm SZ}$, but the lack of understanding of some of the aspects of 
the cluster physics that are introduced in the simulations.

\section{Conclusions.}

We have explored the parameter space of the concordance $\Lambda$CDM model 
using two SZ contributions: one due to the combined KSZ and TSZ effects from  
the unresolved cluster population, the other due to the TSZ from WHIM. 
We have fitted this model to the combined WMAP 7yr and SPT data, at present the best 
data-set publicly available, covering the multipole range $\ell=[2,3000]$.
We have found that a WHIM component with an amplitude of $\sim 20~\mu$K$^2$ 
at $\ell\sim 300$ is compatible with the data. This new WHIM component 
results in a $10$\% decrement in the amplitude of the SZ signal ascribed 
to galaxy clusters, while the cosmological parameters do not change significantly; 
$\Omega_{\rm B}h^2$ is $2$\% higher, and $\sigma_8$ decreases by 
$1$\%. This new fit does not solve the discrepancy between the measured 
amplitude of the cluster power spectrum and the theoretical expectation; this 
would require a considerably lower value of $\sigma_8$. However, 
since $\sigma_8$ is tightly constrained by the data, this  could be an 
indication of improper modelling of cluster evolution,
gas dynamics, and of the resulting pressure profiles, both theoretically 
and numerically. Our analysis shows that a WHIM contribution is compatible 
with WMAP and SPT data and, if this 
contribution is well described by our model, the average properties of the 
WHIM can be measured.  Since the cosmological parameters 
($\sigma_8$, $\Omega_{\rm B}$ and $h$) are determined from the CMB
anisotropy data, measuring the WHIM contribution allows to constrain the 
other two parameters, namely, the cut-off length of the baryons power spectrum, 
parametrized by $T_{\rm IGM}$, and the phase diagram parameter 
$\alpha$ that also determine the amplitude of the WHIM anisotropy. 
For $A_{\rm WHIM}\sim 20\mu K^2$ the Kang et al. (2005) phase diagram 
parameter is constrained to be in the interval $\alpha\simeq [2.5,3.4]$, 
consistent with the estimates derived from simulations.  
The fraction of baryons that would reside in this phase,
between 43\% and 47\%, would be large enough to close 
the baryon census problem summarized in Shull et al. (2012).
These are the first constraints on the 
physical state and abundance of the WHIM presented to date. 

Planck, with its wide coverage of frequencies and angular scales, could help 
to unambiguously measure this WHIM component, separating it from the cluster 
contribution, and to set more stringent constraints
on the WHIM physical properties. The power spectrum 
amplitude scales as $C_\ell^{\rm WHIM}\propto A_{\rm WHIM}G(\nu)^2$.
If $A_{\rm WHIM}=10-15~\mu{\rm K}^2$ at Planck frequencies it will
change from zero at 217~GHz to $\sim 60~\mu{\rm K}^2$ at 40~GHz (decrement) 
and 353~GHz (increment).
This represents a 2\% variation across the observed range.
If foreground contributions, that also change with frequency, are
subtracted down to this level, then the variation of the power spectrum
will be an indication of the WHIM TSZ contribution.

\section*{Acknowledgments}
RGS acknowledges support from the Cosolider Ingenio programme of the 
Spanish Ministerio de Econom\'{\i}a y Competitividad (project 
``Exploring the Physics of Inflation'').
ISV thanks the DAAD for the financial support, grant A/08/73458.
FAB acknowledges financial support from the Spanish
Ministerio de Econom\'{\i}a y Competitividad (grants FIS2009-07238, FIS2012-30926
and CSD 2007-00050). He also thanks the hospitality of
the Leibniz Institute f\"ur Astrophysik.

\bibliographystyle{mn2e}

\pagestyle{plain}

\bsp
\label{lastpage}
\end{document}